\documentclass[iop]{emulateapj}

\usepackage{longtable}
\usepackage{amsmath}
\usepackage{selectp}
\usepackage{mathtools,amssymb}
\usepackage{multirow}
\usepackage[utf8]{inputenc}
\usepackage{nicefrac}
\usepackage{xfrac}
\usepackage{graphicx}
\usepackage{comment}
\usepackage{natbib}

\newcommand{\til}{\raise.35ex\hbox{$\scriptstyle\sim$}}

\begin{document}

\title{A Deep Search for Additional Satellites around the Dwarf Planet Haumea}
\author{Luke D. Burkhart\altaffilmark{1,2}, Darin Ragozzine\altaffilmark{1,3,4}, Michael E. Brown\altaffilmark{5}}
\email{darin.ragozzine@gmail.com}
\altaffiltext{1}{Harvard-Smithsonian Center for Astrophysics, 60 Garden Street, Cambridge, MA 02138, USA}
\altaffiltext{2}{Yale University, Department of Physics, 217 Prospect St, New Haven, CT 06511, USA}
\altaffiltext{3}{University of Florida, 211 Bryant Space Science Center, Gainesville, FL 32611, USA}
\altaffiltext{4}{Florida Institute of Technology, Department of Physics and Space Sciences, 150 West University Boulevard, Melbourne, FL 32901, USA}
\altaffiltext{5}{California Institute of Technology, Division of Geological and Planetary Sciences, MC 150-21,
Pasadena, CA 91125, USA }

\slugcomment{Search for Additional Satellites around Haumea}

\begin{abstract}
Haumea is a dwarf planet with two known satellites, an unusually high spin rate, and a large collisional family, making it one of the most interesting objects in the outer solar system. A fully self-consistent formation scenario responsible for the satellite and family formation is still elusive, but some processes predict the initial formation of many small moons, similar to the small moons recently discovered around Pluto. Deep searches for regular satellites around KBOs are difficult due to observational limitations, but Haumea is one of the few for which sufficient data exist. We analyze Hubble Space Telescope (HST) observations, focusing on a ten-consecutive-orbit sequence obtained in July 2010, to search for new very small satellites. To maximize the search depth, we implement and validate a non-linear shift-and-stack method. No additional satellites of Haumea are found, but by implanting and recovering artificial sources, we characterize our sensitivity. At distances between $\sim$10,000 km and $\sim$350,000 km from Haumea, satellites with radii as small as $\sim$10 km are ruled out, assuming an albedo ($p \simeq 0.7$) similar to Haumea. We also rule out satellites larger than $\gtrsim$40 km in most of the Hill sphere using other HST data. This search method rules out objects similar in size to the small moons of Pluto. By developing clear criteria for determining the number of non-linear rates to use, we find that far fewer shift rates are required ($\sim$35) than might be expected.  The non-linear shift-and-stack method to discover satellites (and other moving transients) is tractable, particularly in the regime where non-linear motion begins to manifest itself.

\end{abstract}


\maketitle

\section{INTRODUCTION}
The dwarf planet Haumea \citep{2006ApJ...639.1238R}, its two moons \citep{2005ApJ...632L..45B,2006ApJ...639L..43B}, and its collisional family \citep{2007Nature..446..296} provide important constraints on the formation of Kuiper belt and outer solar system. This well-studied object is the fastest-rotating large body in the solar system \citep{2006ApJ...639.1238R} with rotational variability in color \citep{2008AJ....135.1749L,2009AJ....137.3404L}, an unexpectedly high density \citep{2014EM&P..111..127L}, and large albedo \citep{2010Natur.465..897E}. It has two moons on dynamically excited orbits \citep[][hereafter RB09]{2009AJ....137.4766R} which have scaled mass ratios and distances similar to the Earth-Moon system. Dynamical, photometric, and spectroscopic observations of objects in the vicinity of Haumea clearly indicate a collisional family of icy fragments with similarly high albedos \citep{2007AJ....134.2160R,2008ApJ...684L.107S,2010A&A...511A..72S}. However, though the expected dispersion velocity of these fragments is of order several hundred meters per second, the observed dispersion is well constrained within $\sim$150 m s$^{-1}$. The apparent lack of high velocity ejecta is confirmed by observational surveys and dynamical studies \citep[e.g.,][]{2012ApJ...749...33F,2012MNRAS.421.1331L,2012Icar..221..106V}, though it is possible that some high velocity ejecta would be unrecognizable dynamically \citep{2011ApJ...733...40M} and/or compositionally \citep{2012A&A...544A.137C,2012AJ....143..146B}.

There is no simple high-probability formation scenario that naturally explains all of these observational constraints: Haumea's rapid near-breakup rotation rate, the two moons on distant and dynamically warm orbits, and a collisional family that is an order of magnitude smaller in velocity dispersion than expected. Though multiple explanations and variations have been proposed \citep[e.g.,][]{2008AJ....136.1079L,2009ApJ...700.1242S,2010ApJ...714.1789L,2011ApJ...733...40M,2012MNRAS.419.2315O,2013AJ....146...89C}, none have adequately and self-consistently explained all of the unique features of this interesting system and its family.

Attempting to place the formation of the Haumea system in context with other similar systems in the Kuiper belt quickly leads to comparisons with Kuiper belt objects (KBOs) of similar sizes, particularly Eris, Pluto, and Makemake. Of these, Pluto is the best understood due to a wealth of observational data and the recent flyby by the New Horizons Mission \citep{2015Sci...350.1815S}. Furthermore, there are similiarities between some of the theories for the formation of Haumea's satellites \citep[e.g.][]{2010ApJ...714.1789L} and for the formation of Pluto's satellites \citep[e.g.][]{2011AJ....141...35C}: both suggest a relatively large impactor with very low incoming velocities that undergo a grazing collision to form a satellite system. With the discovery of a retinue of small satellites exterior to Charon's orbit -- now dubbed Styx, Nix, Kerberos, and Hydra -- there is a renewal in interest in observational constraints on the formation of the Pluto system \citep{2006Natur.439..943W,2011IAUC.9221....1S,2012IAUC.9253....1S,2015Natur.522...45S}. Standard explanations for the formation of Nix and Hydra were already problematic \citep{2006Sci...313.1107W,2008arXiv0802.2951L}, and the characteristics of Styx and Kerberos are even more puzzling \citep[][]{2012CeMDA.114..341P,2014AJ....147....8K,2014arXiv1407.1059C}. For example, in the current orbital configuration, the dynamical stability of Styx requires that Charon's eccentricity at its present semi-major axis was never above $\sim$0.035, using the circumbinary stability criterion of \citet[][see Equation 3]{1999AJ....117..621H}. Thus, the discovery of Styx combined with dynamical stability immediately precludes some of the more extreme proposed orbital histories of \citet{2014Icar..233..242C} if Styx formed concurrently with Charon \citep{2014arXiv1407.1059C}. Long-term dynamical stability can also place some of the best constraints on the masses of these small moons \citep{2012ApJ...755...17Y,2013AJ....146...89C,2015arXiv150505933P,2015Natur.522...45S}.

The discovery of small moons around Pluto and their ability to add constraints to the understanding of this system suggests that all asteroid and KBO binaries and triples be searched for additional moons. We recommend the continuation of this standard practice, even when an initial companion is identified. For KBOs, satellite searches are observationally difficult for multiple reasons. First, acquiring data of sufficient depth and resolution to identify faint moons of faint KBOs usually requires a considerable amount of time at the best telescopes in the world, such as the \emph{Hubble Space Telescope} (HST) or 8-10 meter class telescopes with Laser Guide Star Adaptive Optics. The only KBOs with large amounts of continuous high-quality data are Pluto and Haumea.

Second, the discovery of small moons can be frustrated by their \emph{a priori} unknown satellite orbital motion during long exposures. Faint, fast-moving moons can then evade detection even with the best data, using standard analysis methods. Therefore, an enhanced methodology to search for faint moving moons is required.

In an attempt to better understand the formation of the Haumea system, we use a large set of consecutive HST observations to perform a search for very small moons around Haumea similar to those discovered around Pluto ($\S$2). To search for faint, fast-moving moons well below the single-exposure limit, we implemented the non-linear shift-and-stack method proposed by \citet[][hereafter PK10]{2010PASP..122..549P} for the discovery of KBOs ($\S$3). Adapted to the problem of finding additional satellites, this method was both efficient and effective. Though no additional satellites of Haumea were detected ($\S$4), with careful characterization of this null detection, we set strong limits on the size and location of possible undiscovered moons ($\S$5) and discuss the implications for understanding of Haumea's satellite system ($\S$6).

\section{OBSERVATIONS}

In determining the ideal set of observations for a deep satellite search, a balance must be struck between including the largest number of observations and considering the motion of putative satellites during the total observational baseline. The standard stacking method of adding images that have been co-registered to the position of the primary to enhance sensitivity to faint satellites is limited to observational arcs where the satellite's relative position remains within a region not much larger than $\sim$1 Point Spread Function (PSF) Full Width at Half Maximum (PSF FWHM). Our use of non-linear shift-and-stack can mitigate this problem significantly and allows us to perform a sensitive search on longer timescales. In particular, our HST Program 12243 observed with a wide filter for 10 consecutive orbits and is an excellent dataset for a deep satellite search; with the technique discussed below, we can search these observations even for close-in satellites which traverse a significant fraction of an orbit during the 15-hour baseline, corresponding to several PSF widths. These observations are our main focus as they are clearly the best for a deep satellite search ($\S$2.1); however, we also inspected other observations for the additional satellites of Haumea ($\S$2.2).

\subsection{HST 12243: 10 Orbits in July 2010}

HST Program 12243 obtained 10 orbits of observations over the course of $\sim$15 hours in July 2010. This program used the Wide Field Camera 3 (WFC3) UVIS imager with the F350LP (long-pass) filter. The primary goal of these observations was the the detection of a mutual event between the inner moon Namaka and Haumea (RB09). In order to produce high cadence time series photometry of the proposed mutual event (and to avoid saturation), exposures were limited to $\sim$45 seconds. To prevent a costly memory buffer download, only a 512x512 subarray of the full WFC3 camera was used with a field of view of $\sim$20.5 arcseconds. HST tracked at Haumea's rate of motion (except for controlled dithering) to maintain its position near the center of the field of view throughout the observations. The geocentric distance to Haumea at time of observations was 50.85 AU. At this distance, 1" corresponds to 36900 km, 1 WFC3 pixel (.04 arcseconds) corresponds to 1475 km,  and the entire subarray field of view corresponds to $\sim$750000 km.

Parts of the last two orbits were affected by the South Atlantic Anomaly. This caused a portion of the these orbits to lose data entirely, and another portion was severely affected with cosmic rays and loss of fine pointing precision. The most offensive frames were discarded for the purpose of the satellite search, leaving 260 individual exposures.

The center of Haumea was identified by eye in combination with a 2-d Gaussian fitting routine. With these well-determined preliminary Haumea locations, all images were co-registered to Haumea's position. In this Haumea-centric frame, cosmic rays and hot pixels were identified by significant changes in brightness at a particular position using robust median absolute deviation filters. A detailed and extensive image-by-image investigation of cosmic rays by eye confirmed that this method was very accurate at identifying cosmic rays and other anomalies. Furthermore, the automatic routine did not flag the known objects (Haumea and its two moons: Hi'iaka and Namaka) as cosmic rays, nor were any other specific localized regions identified for consistent masking (i.e., putative additional satellites were not removed).

TinyTim software\footnote{\texttt{http://www.stsci.edu/hst/observatory/focus/TinyTim}} was used to generate local point-spread function (PSF) models. As in RB09, these PSF models were then fit to the three known objects using standard $\chi^2$ minimization techniques \citep{2009ASPC..411..251M}. This identified the best-fit locations and heights of the scaled PSFs, with bad pixels masked and thus not included in the $\chi^2$ calculation. The astrometric positions of the known satellites relative to Haumea seen by this method were in clear accord with their projected orbital motion from RB09. Despite Namaka's nearness to Haumea (which was purposely chosen, as the goal of the observations was to observe a mutual event), it is easily distinguishable in the first few orbits.

The best fit PSFs are visually inspected and found to be good fits for all images. These PSFs are then subtracted, removing a large portion of the three signals, but leaving non-negligible residuals; these residuals are caused by imperfect PSFs (note that Haumea may be marginally resolved in some images) and standard Poisson noise. Using the updated best-fit centers of Haumea, these PSF-subtracted images are re-coregistered to Haumea's best-fit location (though the additional shifts from the preliminary 2-d Gaussian centers are very small). As described below, these same PSF-subtracted images are used to perform the non-linear shift-and-stack.

\begin{figure}
\centering
\plotone{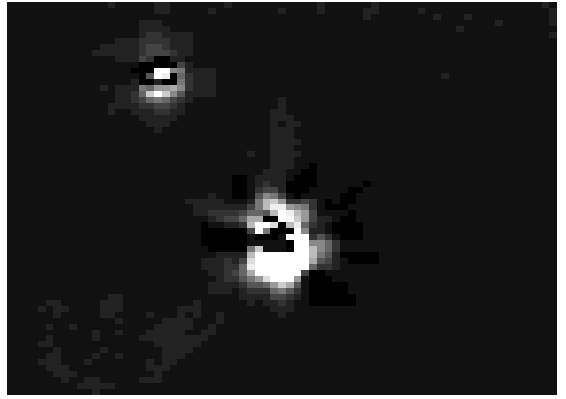}
    \caption{A portion of the median stack of 260 images from 10 orbits of HST WFC3 data (Program 12243). Individual images are co-registered to be stationary in the Haumea-centric frame, with best-fit TinyTim PSFs for Haumea, Hi'iaka and Namaka subtracted. The brightness has been stretched significantly to highlight the residuals. These residuals will limit sensitivity near Haumea, but the diffraction spikes and the majority of the PSFs have been removed. Above and to the left of Haumea lie the residuals from Hi'iaka which is 1.23" away (45600 km projected distance) in this stack. Vertical darker columns are due to minor uncorrected pixel sensitivity. The image is aligned so that Astronomical North is up. }
	\label{stationary}
\end{figure}

Moons with negligble motion relative to Haumea can be identified in this image by stacking all the observations in the Haumea-centered reference frame (including fractional pixel shifts implemented by IDL's \texttt{fshift}). Throughout this investigation, we create stacks by performing a pixel-by-pixel median of the images, which is less sensitive to cosmic rays, bad pixels, and errors in the PSF subtraction of the known bodies. This results in a small decrease in sensitivity, as, assuming white noise, the noise level grows a factor of $\sqrt{\frac{\pi}{2}}$ faster when using the median over the mean. This will correspond to only a $\sim 20\%$ difference in brightness sensitivity, or a $\sim 10\%$ difference in radius, and we find this acceptable so that the other effects mentioned above can be mitigated. A portion of this stacked image around Haumea is shown in Figure \ref{stationary}. Detailed investigation of this deep stack by eye by each co-author yielded no clear satellite candidates. The median stacks were also automatically searched using the IDL routine \texttt{find}, which uses a convolution filter with FWHM of $1.6$ pixels to identify positive brightness anomalies. Detections with SNR of 5 or greater were examined; none were found that were consistent with an additional satellite (e.g., having PSF-like shape). Scaling from the SNR of Haumea and using a more conservative detection limit SNR of 10, this places a limit on non-moving satellites as fainter than about $V\simeq27.1$, corresponding to Haumea satellites with radii less than about $8$ km (see $\S$5).

\subsection{Other Observations}

HST has observed Haumea during many programs for multiple reasons. Program 11518 was proposed to obtain astrometry of both moons and is 5 independent orbits of observations spread over 2 weeks (RB09). Although it would be interesting to investigate the possibility of combining these data in a long-baseline non-linear shift-and-stack, given the existence of other more sensitive datasets, we investigated only the single-orbit median stacked images. Motion during a single 45-minute HST orbit is small compared to the PSF width, even for the shortest satellite orbital periods.

HST Program 11971 was 5 consecutive orbits and HST Program 12004 was 7 consecutive orbits, both attempts to observe the last satellite-satellite mutual events. The latter program was within a few weeks of the HST 4th Servicing Mission but was still executed. Unfortunately, for 6.5 of the 7 orbits, the STIS shutter was closed and no on-sky data were taken.

For the 5-orbit Wide Field Planetary Camera 2 observation of Program 11971, we median-stacked images centered on Haumea and searched for additional sources by eye and using IDL's \texttt{find} as described above. We investigated stacks of individual orbits and of the entire 5-orbit sequence and found no sources consistent with additional satellites. Though the non-linear shift-and-stack method below could fruitfully be applied to these observations, the WFC3 observations are considerably deeper and we opted to focus on our best dataset.

Finally, we obtained some long-duration ($\sim$5 hours) observations of Haumea using the Laser Guide Star Adaptive Optics system at the Keck Observatory. Co-registered stacks of this data also showed no clear additional satellites, though the known satellites were very easily detected.

\section{METHODS}

For the detection of faint bodies, with signal-to-noise ratio (SNR) per image of $\lesssim$5, a useful approach is the co-addition (``stacking'') of multiple images. With the 260 images in our dataset, this method can increase the SNR by $\sim$$\sqrt{\frac{2}{\pi}}\sqrt{260} \approx 13$, thereby searching for satellites with radii $\sim\sqrt{13}\approx 3.6$ times as small as could be detected in a single image.

If the object does not remain apparently stationary (within $\lesssim$1 FWHM) over the course of the observation, the simple co-addition will result in insufficient overlap between images to yield the expected increase in SNR. If the motion of the object is known, images can be first shifted to compensate for this motion, and the images added with the object localized regaining nearly the full sensitivity: this is the meaning of ``shift-and-stack''. Linear searches with shift-and-stack have been used to discover satellites in the past \citep{2004Icar..169..474K,2004Natur.430..865H, 2013DPS....4520601S} although these searches did not need to use the non-linear shift-and-stack method we employ below.

In a situation where the motion is unknown, such as a broad search for KBOs, a large set of possible paths on the sky can be considered, and each path independently used as the basis for a shift-and-stack, as described by PK10. A composite image results from each proposed orbital path (which we call a ``sky track''), and each stack can be searched for faint satellites which emerge from the noise due to shifting the image accurately enough to (mostly) compensate for its motion.

To minimize statistical false-positives and to increase computational tractability, it is important to identify a near-minimal number of sky tracks that will faithfully reproduce all the possible motions without performing redundant searches. PK10 suggested an algorithm for identifying the most important non-redundant set of sky tracks, which we fruitfully employ: generate a large number of random sky tracks based on the full range of expected motion (within desired search parameters) and then remove tracks that are similar to one another. We have adapted this technique for our search.

There is a distinction between a general KBO search and a satellite search, which, it is important to note, is largely ignored in the method presented here. This distinction is that, in a broad KBO search, a sky track could be valid for any part of an image; that is, there is little correlation between position and motion. This is not the case for a satellite orbiting a given primary, in which a specific motion only applies to a small spatial region. The more highly curved tracks are, the more specific to a particular region they are --- a curved orbital arc translated to the other side of Haumea would not make physical sense. The method described below involves shifting and stacking the entirety of each image, and searching the whole of the composite image, when in fact the track upon which the shift-and-stack is based applies to only a small subset of each image. In addition to the computational cost of shifting and searching larger images than is necessary, this overuse of the images could potentially result in an increase in statistical false-positives. However, neither of these effects manifest in a noticeable way --- neither computation time nor an abundance of false-positives limit our search method. This suggests that we are near the optimal minimum number of sky tracks searched, or have at least reached an acceptably small number.

Discussed in greater detail below, an overview of our search algorithm is as follows:
\begin{enumerate}
\item Generate a large bank of physically reasonable putative sky tracks by randomly selecting from plausible Keplerian satellite orbital parameters.
\item Fit each sky track with non-linear polynomials in time (shift rates). If the shift rates for two distinct sky tracks are similar enough (quantified below), discard one.
\item Continue searching for sky tracks until a nearly-complete non-redundant set is identified.
\item For each track, create a composite image. This is done by overlaying the dataset (in our case, 260 images) upon itself, with the images shifted by the appropriate shift rates such that an object on that track will appear in the same place in each image. Co-add the images into one composite.
\item Search each composite image for satellite candidate sources.
\end{enumerate}
The use of non-linear polynomial fits allows the shift rates to more accurately capture curved orbits than simple linear fits. For the motion of even the fastest detectable Haumean satellites over the timescale of our observations, we find that quadratic fits to the x and y positions are always sufficient. Note that the polynomial fits are included for convenience in describing the sky tracks; the actual positions of a putative satellite could be used, but the difference between the actual positions and the best-fit quadratic approximate was negligible. Including non-linear rates is often expected to greatly expand the number of dissimilar shift rates to the point of computational impracticality, but we find that an appropriate criterion for similarity of shift rates easily permits the inclusion of quadratic rates.

\subsection{Generation of Sky Tracks}

In a typical shift-and-stack search for KBOs, the putative sky tracks are selected from a grid of the six degrees of freedom needed to describe an object in a Keplerian orbit \citep[PK10, ][]{2004AJ....128.1364B}. For the purposes of a KBO satellite, particularly that of a primary with other known satellites, it is convenient to instead sample the space of Keplerian orbital elements relative to the primary: semi-major axis ($a$), eccentricity ($e$), inclination ($i$), longitude of the ascending node ($\Omega$), argument of periapse ($\omega$), and mean anamoly at epoch ($M$). Sampling in this space allows for direct control over the types of orbits that are searched, making it straightforward to exclude unphysical motions. In our case, we also benefit from a well-known mass of the primary; if this is not known, a variety of plausible values could be sampled for the generation of sky tracks. For this search, $a$ and $e$ were randomly sampled from orbits with semi-major axes between 5310 and 368000 km and eccentricities less than 0.5, while the orbital angles $i$, $\Omega$, $\omega$ and $M$ were allowed to assume any value.  All parameters were chosen from the sample space uniformly, with the exception of $a$, which was sampled on a log scale to increase the likelihood of sampling an orbit in the regime of fast-moving satellites.

The lower bound on $a$ is a constraint imposed by our sensitivity of detection. This limit corresponds to 3.75 pixels (15 milliarcseconds) on the WFC3, at which distance from the center of Haumea the subtraction noise is considerable enough to make reliable detections difficult (see Figure \ref{stationary}). The upper bound on $a$ is set much larger than the semi-major axes needed to shift images (as opposed to investigation of the unshifted stack). At a distance where the satellite's maximum velocity would cause it to travel less than one PSF FWHM over the course of the 15 hour observational period (here $\sim$27 m s$^{-1}$), shift-and-stack is unnecessary, giving the upper limit of $a \simeq 150,000$ in our search. This semi-major axis is $\sim$3 times the semi-major axis of Hi'iaka, whose motion in these frames is detectable, but $\lesssim$0.5 pixels. For the upper limit on $a$, we doubled this number to be conservative. \label{notmoving}

Much of this orbital parameter space can be excluded on physical grounds, reducing the number of shift rates necessary to well-sample the space. Any putative orbit which crossed paths with the known satellites was rejected, as was any orbit with periapsis less than $3000$ km. These weak restrictions on orbital elements did not appreciably affect the selection of shift-and-stack parameters and additional tests (described below) show that we are sensitive to objects on practically any orbit with semi-major axis $\gtrsim$10000 km.

\subsection{Non-linear Fitting and Shift Rate Similarity}

Having created a bank of physically plausible orbits, we then generate a set of shift rates with which to create composite images to search for satellites. Orbital parameters were converted into sky coordinates relative to Haumea, right ascension ($\Delta$RA) and declination ($\Delta$Dec), for each image, as described in RB09. We assumed an instantaneous Keplerian orbit for the position of the satellite as this is an excellent approximation over the course of our observations. In our case, orbital acceleration was quite important, as we desired to search orbital periods down to $\sim$40 hours, of which the 15-hour observational arc is a sizable fraction. Therefore, the sky positions $\Delta$RA and $\Delta$Dec were fit with quadratic polynomials in time, which we found were sufficient to accurately describe the non-linear motion in every case.

In order to minimize the number of sky tracks, we eliminated tracks which were similar to one another, as suggested by PK10. To determine if two tracks were similar, we focused on the final requirement that the shift rates localize the flux of satellite so that it can be identified in the stacked image. If the flux of a satellite traveling along the second orbit would be well-localized by the shift rates of the first, then there would be considerable overlap of the flux between images when shifted according to the rates of the first. This criterion can be quantified by calculating the overlap fraction between two shift-and-stack rates using the reasonable assumption that the WFC3 PSF is nearly Gaussian, with FWHM of .067 arcseconds ($\approx$1.7 pixels). For a pair of shift rates, the overlap was defined for each image in the dataset as the integral of the product of two such Gaussians separated by the difference in the two rates ($\Delta$RA and $\Delta$Dec) at the time of that image. We call this the overlap between two orbits as it is calculated from the product of two overlapping PSFs, but it is distinct from the concept of the overlap in co-added images. If the median overlap (normalized to 1 for perfect coincidence) was greater than a pre-specified threshold, it was considered that a sufficient fraction of the flux of the proposed satellite would have been collected by the stack of an existing sky track, and the new track was rejected as unnecessary.

The goal is to build up a bank of sky tracks known to be mutually distinct. After accepting the first track into the bank, each subsequent track was compared to the previously selected tracks in the bank using the above overlap criterion. We experimented with different overlap threshold criteria and found the overall results mostly insensitive to the specific value chosen. In general, we required a median overlap of less than 0.7 with each previously accepted shift-and-stack tracks to accept the proposed track as distinct enough to add to the bank.

By drawing from a large set of orbits covering the desired search space, this method efficiently builds a bank of mutually distinct shift rates that are also the most relevant (PK10). However, unlike a grid search, random orbital draws can continue indefinitely. Thus, we also require a ``stopping criterion'' to decide when the bank is large enough for practical use. To determine the upper limit on the number of necessary shift rates, we noted that the sample space saturates quickly; that is, the rate of acceptance drops off drastically after 10-15 shift rates are chosen. Consequently, the number of orbits rejected between successive accepted rates grows very quickly. Our criterion for a dense sampling was that the number of rejected rates between successive accepted rates was at least equal to the total number of rates rejected so far. Put another way, the selection was stopped when the acceptance of each new shift rate required doubling the number of sampled orbits, which typically occurred after testing hundreds of thousands of random orbits. This is an exponentially slow process, which suggests that once past this threshold, we have reached the limits of our rate selection method. Practically speaking, we found that this stopping criterion still generated enough shift rates to recover injected sources with a complete variety of orbits.

\begin{figure}
\centering
\plotone{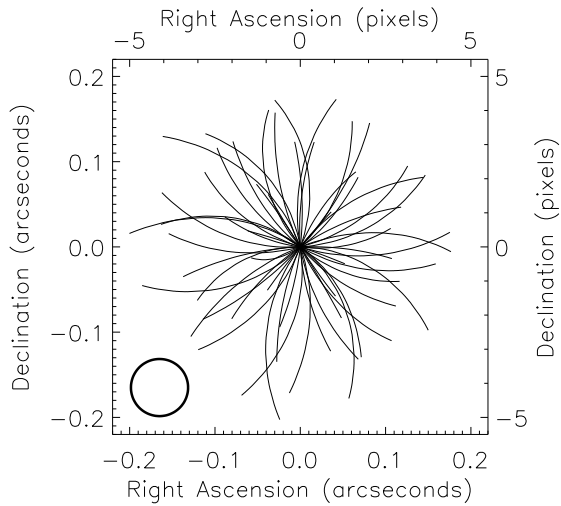}
    \caption{The non-linear shift-and-stack rates. Arcs show the displacement of images (relative to position at the middle image) over the course of the 15 hour observation. Each arc represents a different shift rate or ``sky track.'' Horizontal and vertical axes show differential right ascension ($\Delta$RA) and declination ($\Delta$Dec) in arcseconds and pixels (1 pixel = 0.04 arcseconds = 1475 km). The circle at bottom left has diameter of .067 arcseconds, the FWHM of WFC3's PSF. Following the method suggested by PK10, we generate random non-linear shift rates out of Keplerian orbital elements. We reject as duplicates rates for which the overlapping PSFs would catch at least 70\% of the flux if moving at the same rate as an existing orbit (see $\S$3.2). This method requires only $\sim$35 non-linear rates to cover the vast majority of parameter space. The ``sky tracks'' associated with these rates are mostly symmetric about the origin as seen above, with slight asymmetries arising from the projection of eccentric orbits into the skyplane, and the variation in orbital speed throughout an orbit. Almost all rates are substantially quadratic, which shows the importance of the non-linear approach. As can be seen, the use of quadratic shift rates allowed us to probe the region near Haumea where satellites would execute sizable fractions of an orbit during the 15-hour observation. Implantation of artificial sources on orbits randomly drawn from the same Keplerian elements showed an excellent recovery rate (see Figures \ref{avsb} and \ref{svsb}).}
	\label{tracks}
\end{figure}

Together with the above ranges for satellite orbital parameters, this method yielded only $\sim$35 sufficiently distinct orbits over which to search. Considering the size of the parameter space (linear and quadratic terms for shifts in both x and y directions), it might seem surprising that so small a set of potential orbits spans the space. However, a large portion of the space consists of short, almost entirely linear tracks, where the quadratic corrections are of limited importance. The only strongly quadratic orbits are very near to Haumea, which also have large linear rates. In other words, there are strong correlations between the allowed linear and quadratic coefficients of physically-plausible tracks. The result is a relatively small number of non-linear shift rates that efficiently cover the desired search space (PK10); these tracks are illustrated in Figure \ref{tracks}. Contrasted with the orbital element motivated sampling presented here, the number of shift rates in the case of a quadratic sky motion grid search would have been much larger.

\subsection{Creation of Composite Image}

With our bank of non-degenerate sky tracks, we can now perform the non-linear shift-and-stack procedure. Each track corresponds to a specific set of $\Delta$RA and $\Delta$Dec values of a putative satellite relative to Haumea. We used \texttt{adxy}, a routine from the IDL Astro Library which uses astrometric data from the image headers, to convert these sky coordinates into on-image pixel positions, thus yielding the desired pixel shifts.

The prepared images were shifted (including fractional pixel shifts implemented by IDL's \texttt{fshift}) and stacked using the pixel-by-pixel median of the images as described above. In preparation for the automated search, many images were investigated in detail by eye.

\subsection{Sensitivity}

To test the sensitivity of this method, artificial sources were implanted into the images with a range of random brightnesses. Their positions and rates of motion were determined by orbits randomly drawn from the same space mentioned above (but without restriction of non-crossing orbits with the known moons). The implants were generated by scaling from the actual PSF of Haumea (when brightest). This source was implanted into the images at the pixel positions corresponding to the randomly-chosen orbit. A subimage of 200 x 200 pixels was used for the search: the outer reaches of this subimage have objects that are practically not moving (see $\S$\ref{notmoving}) and any object beyond this region would have the same detectability threshold as stationary objects.

This was done for a large number of orbits, with a new set of images created for each. Stacks were generated in the exact same manner as the real images, with the same $\sim$35 shift rates making new median stacks for each new set of images. These stacks were inspected using the same automated search routine (IDL's \texttt{find}). To distingush detections of implanted sources from the detection of the three known bodies, we examined the output of the search for the sets of stacks with no sources implanted. All detections here were due to known bodies, and the positions were used to establish a mask with three regions, one for each known body, to reject detections that were not due to implanted sources. In this way, detections could be automatically classified as a recovery of an implant or as a false positive due to the known bodies. These automated classifications were extensively verified with an investigation that included searching by eye and found to be very robust.

Due to the application of a threshhold SNR by the \texttt{find} routine, objects in the vicinity of Haumea, while still far enough and bright enought to be seen by eye, may be rejected by the routine itself (not our masks). The presence of the primary nearby leads to an artificially high computed background noise level, which reduces the computed SNR significantly, causing the object to appear below threshhold. Any stacks with sources at risk of being left undetected due to this effect were searched by eye by multiple coauthors, and any that were detected in this processes were considered to be recovered for the purposes of our results, shown below.

The success rates of finding implanted objects places constraints on additional satellites of Haumea: any recovered implanted source represents a satellite that we can say with reasonable certainty is not present in the Haumea system.

\section{RESULTS}
The implantation and successful recovery of faint moving sources clearly indicated the effectiveness of our non-linear shift-and-stack method. Nevertheless, we did not detect any additional satellites around Haumea and no candidate satellites were found that were worthy of additional investigation.

\begin{figure}
\centering
\plotone{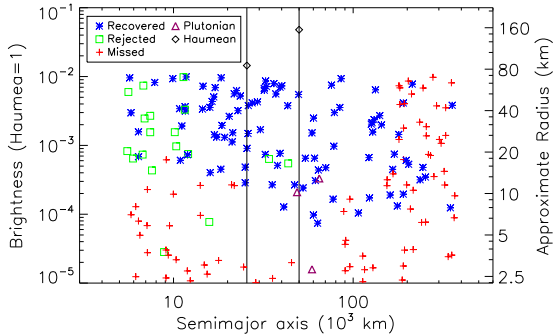}
	\caption{(Color online) Results from the sensitivity survey. The figure shows implanted sources that were either recovered (blue stars), not recovered (red crosses), or recovered but rejected due to confusion with existing sources (green squares). The horizontal axis is the semi-major axis of implanted objects in thousands of kilometers. The left-hand vertical axis is brightness relative to Haumea (when brightest). The right-hand vertical axis is radius of a spherical satellite assuming an albedo ($\sim$0.7) similar to Haumea. Diamonds represent know satellites Namaka and Hi'iaka; the vertical lines are guides to the eye at their respective semi-major axes. Purple triangles represent the moons of Pluto --- Nix, Hydra and Kerberos --- according to brightness relative to the primary. (The smallest moon Styx, with brightness approximately $6\times 10^{-6}$ that of Pluto, is below the range of brightness represented on this figure.) Because of differences in geocentric distance and albedo, the approximate radius does not directly apply to these three points. Figure \ref{svsb} is similar but shows distance in projected separation instead of semi-major axis. Bounds on brightness and semi-major axis of were chosen as described in $\S$3.1. The unrecovered implantations at semi-major axis $\gtrsim 200\times 10^3$ km are not found because their distance from Haumea often places them outside the subimages searched. This figure shows that satellites with radii as low as $\sim$8 km would be detectable in much of the space searched, and that our lower detection limit on semimajor axis is limited by the properties of the dataset, not by the sensitivity of the non-linear shift-and-stack technique. Nix and Hydra-like objects would be detected around Haumea, while Styx and Kerberos-like objects would still be too faint, mostly due to Haumea's further distance (50 AU compared to Pluto's 30 AU).  }
	\label{avsb}
\end{figure}

\begin{figure}
\centering
\plotone{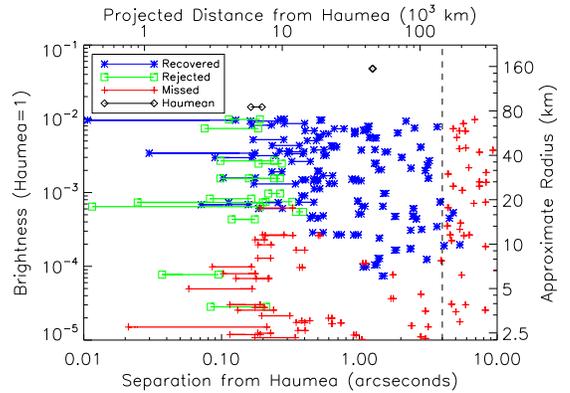}
	\caption{(Color online) Results from the sensitivity survey. The lower horizontal axis is the sky-planet projected separation from Haumea in arcseconds, while the upper axis gives approximate projected distance from Haumea in thousands of kilometers. The vertical axes are brightness and radius of implanted sources, as described in the caption to Figure \ref{avsb}. Symbols connected by horizontal lines show the maximum and minimum apparent distance from Haumea of the implanted object during the 15-hour "observation." As in Figure \ref{avsb}, implanted sources were either recovered (blue stars), not recovered (red crosses), or recovered but rejected due to confusion with existing sources (green squares). Note that implantations at separation greater than 4 arcseconds are unrecovered because they fall outside the region of the $\sim$4" subimages that were searched (see $\S$5.3). The vertical dashed line is a guide to the eye for this rough cutoff. No sources were implanted at separations larger than 10 arcseconds, corresponding to the upper limit on semi-major axis shown in Figure \ref{avsb}. Diamonds represent Hi'iaka and Namaka as they appear in the observaion; Namaka's separation is only given for the first four orbits where its presence is measured reliably enough for precise astrometry; as these observations were designed to catch Namaka in a mutual event, its projected separation would approach very low values if all ten orbits were included.}
	\label{svsb}
\end{figure}

A careful characterization of this null result is able to place strong limits on the brightness and separation of undiscovered Haumean satellites. These limits are summarized by Figures \ref{avsb} and \ref{svsb}, which show the results of our search for each implanted source. The source is either recovered, rejected (for being too close to one of the three known objects, usually Haumea), or ``missed'' because was too faint to be detected, or because it fell off the 200 x 200 subimage that was searched. Note that the ``rejected'' category is primarily composed of objects that we not clearly detected by the automated routine, but were detected in a blind search by eye by multiple coauthors; these consistent entirely of objects that are $\lesssim$0.2" (5 pixels) from Haumea. Figure \ref{avsb} plots semi-major axis against brightness of the sources as a fraction of that of Haumea, while Figure \ref{svsb} shows the projected distance (in arcseconds) of the moving sources versus the brightness. Assuming the same albedo ($p \simeq 0.7$) as Haumea, the relative brightness corresponds to the radius of a spherical satellite, which is also shown.

\section{DISCUSSION}

The constraints on undiscovered Haumean satellites can be divided into three categories based on orbital semi-major axis: close-in satellites ($a$ $\lesssim$ 10000 km), intermediate satellites (10000 km $\lesssim$ $a$ $\lesssim$ 350000 km), and distant satellites ($a$ $\gtrsim$ 350000 km).

\subsection{Limits on Close-in Satellites}

At a semi-major axis of $\sim$10000 km, the maximum separation of a satellite from Haumea would be 6.9 WFC3 pixels. Within 7 pixels ($\lesssim 4$ PSF FWHM) of Haumea, it is very difficult to recover objects due to imperfect subtraction of Haumea's PSF. It is possible that an empirical PSF subtraction would perform better for recovering very close-in satellites, but we do not consider such an approach here. As can be seen in Figure \ref{svsb}, there is the expected anti-correlation between the brightness of an object that can be recovered and the separation from Haumea: close in, only brighter objects can be found.

However, there are dynamical reasons to expect that this region is nearly devoid of satellites. Due to Haumea's highly triaxial shape, the orbital region near Haumea is strongly perturbed and long-term stable orbits are difficult to maintain. According to \citet[][]{1994Icar..110..225S}, periods less than about 10 times the spin period are unlikely to be stable due to primary-spin-satellite-orbit resonances. In Haumea's case, this is exacerbated by the additional effects of tidal evolution and other dynamically excited satellites \citep[][]{1999AJ....117..603C,2013AJ....146...89C,2014arXiv1407.1059C}. An orbital period that is 10 times the spin period corresponds to a semi-major axis of about 5000 km (about 5 times the long-axis radius of Haumea). While about twice as distant as the Roche radius, for long-term dynamical stability, we consider this the inner limit.

Even if satellites were originally found in such short orbits, it is possible that long-term tidal evolution would have moved them to a more detectable distance. A detailed analysis by \citet{2013AJ....146...89C} calls into question the idea first proposed that the satellites tidally evolved outwards from orbits near the Roche lobe. While extensive tidal evolution might not have taken place, it is worth noting that scaling the tidal evolution from the properties of the other satellites \citep{2005ApJ...632L..45B}, indicates that even for the smallest satellites we could have detected (which evolve the shortest distance due to tides), tidal evolution would have placed them near or beyond the $\sim$5000 km detection threshold.

There remains a range of semi-major axes from 5000-10000 km that could potentially harbor very small undetected satellites which would be somewhat protected from dynamical and tidal instability. By lying well within Haumea's PSF, these satellites also generally evade detection. Furthermore, some satellites would not have been detected if they had an orbital phase placing them at undetectably small distances (although this is mitigated somewhat by observations at a variety of times).
Overall, it is difficult to hide stable inner ($a \lesssim$10000 km) Haumean satellites with radii $\gtrsim$30 km.

\subsection{Intermediate Satellites}

At semi-major axes between about 10000 and 350000 km lies the region of  satellites near where the other two moons are detected (at semi-major axes of 25600 km for Namaka and 49900 km for Hi'iaka, RB09). At this distance, contamination from Haumea is negligible and the main limitation to detecting satellites is insufficient SNR or falling beyond the edge of the image. By using the non-linear shift-and-stack we maximize the search depth, particularly closer to Haumea.

The search depth can be reported as relative brightness (in magnitudes and flux) and as the radius of a spherical satellite assuming the same albedo as Haumea. As is usual for such deep searches, the recovery rate is a function of magnitude (Figure \ref{avsb},\ref{svsb}). We reliably detect satellites at -9.2 magnitudes (0.0002 relative brightness, radius of 10 km), our recovery rate is roughly 50\% at -10 magnitudes (0.0001 relative brightness, radius of 8 km), and our best case recovery is at -10.4 magnitudes (0.00007 relative brightness, radius of 6 km). Following typical practice, we summarize the recovery depth using the 50\% recovery rate. Note that it is possible that the albedo of the satellites is even higher; using Haumea family member 2002 TX300's measured albedo of 0.9 \citep{2010Natur.465..897E} instead of Haumea's presumed 0.7 albedo \citep{2014EM&P..111..127L} would imply a radius detection threshold of only $\sim$7 km (or $\sim$5 km in the best case).

While close approaches to Hi'iaka and Namaka as projected on the sky would result in a missed detection for faint objects, this is generally unlikely (even for orbits coplanar with the known satellites which are near edge-on, RB09). Close approaches to Hi'iaka and Namaka are negligibly unlikely to happen at more than one epoch\footnote{Unlike irregular satellites of the giant planets, long-term tidal stability precludes Hi'iaka or Namaka from being binaries themselves.}, thus any missed detection would be mitigated for moderately bright objects by the non-detection of satellites in other datasets. Thus, we expect that this region of the Haumean system does not contain undiscovered satellites larger than $\sim$8 km in radius.

Our results compare favorably with the current state of knowledge regarding the small satellites of Pluto. From the New Horizons flyby, we now have detailed knowledge of the albedoes (about 0.5) and sizes of the small satellites: $\sim$10 km for Styx and Kerberos and $\sim$40 km for Nix and Hydra \citep{2015Sci...350.1815S}. As Figure \ref{avsb} shows, we predict that a satellite of apparent magnitude relative to Haumea similar to that of Hydra or Nix around Pluto (-8.7 and -9.2 magnitudes respectively) would fall above our detection limit. With the higher expected albedo (0.7) of Haumean satellites, we would have detected objects as large as Styx and Kerberos. We conclude that Haumea very likely does not contain small satellites similar to Pluto's.

\begin{deluxetable*}{llrrrrrc}
\label{magnitudes}
\tabletypesize{\footnotesize}
\tablewidth{0pt}
\tablecaption{Summary of Estimated Properties of Dwarf Planet Satellites}
\tablehead{\colhead{Object} & \colhead{Satellite} & \colhead{Relative Brightness} &
\colhead{$H_{sat}$\tablenotemark{a}} & \colhead{$V_{sat}$\tablenotemark{a}} &
\colhead{Radius\tablenotemark{b}} &
\colhead{$a$\tablenotemark{c}} &
\colhead{Ref}\\
 & & (magnitudes) & & & (km) & ($10^3$ km) &}

\startdata
Haumea & Hi'iaka & $-3.3$ & $3.4$ & $20.5$ & 200 &  $50$ & 1 \\
Haumea & Namaka & $-4.6$ & $4.7$ & $21.8$ & 150 & $26$ & 1\\
Haumea & ``close'' upper limit & $-6.7$ & $6.8$ & $23.9$ & 30 & $\lesssim$10 & 2\\
Haumea & ``intermediate'' upper limit & $-10.0$ & $10.1$ & $27.6$ & 8  & 10-350 & 2\\
Haumea & ``distant'' upper limit & $-6.2$ & $6.3$ & $23.4$ & 40 & $\gtrsim$350 & 2\\
\hline
Pluto & Charon & $-2.6$ & 1.9 & 16.6 & 350 & $20$ & 3 \\
Pluto & Hydra & $-8.7$ & 8.0 & 22.7 & 41 & $64$ & 3 \\
Pluto & Nix & $-9.2$ & 8.5 & 23.2 & 35 & $49$ & 3 \\
Pluto & Kerberos & $-12$ & 11 & 26 & 12 & $59$ & 4 \\
Pluto & Styx & $-13$ & 12 & 27 & 11 & $42$ & 5 \\
\hline
Eris & Dysnomia & $-6.7$ & 5.5 & 25.4 & 60 & $37$ & 6 \\
Eris & ``close'' upper-limit & $-5.8$ & 4.6 & 24.5 & 80 & $\gtrsim$18 & 6 \\
Eris & ``distant'' upper-limit & $-8.2$ & 7.0 & 26.9 & 30 & $\gtrsim$37 & 6 \\
\hline
Makemake & S/2015 (136472) 1 & $-7.8$ & 7.4 & 24.7 & 25 & $\sim$100 & 8\\
Makemake & upper-limit &$-10$ & 9.6 & 26.9 & 8 & $\gtrsim$30 & 7,8 \\

\enddata

\tablecomments{Magnitudes and semi-major axes of bodies in KBO systems. The relative magnitude of the faintest detectable bodies in our search is -10, comparable to that of Hydra and Nix. For Eris and Makemake, values are more approximate and/or interpolated from published estimates. We do not list the large number of KBO binaries \citep[e.g.][]{2008ssbn.book..345N} or KBO triple 1999TC36 \citep{2010Icar..207..978B} since the formation of these systems appears to be
distinct from processes associated with dwarf planets. In particular, these binaries tend to be nearly equal brightness without known small additional companions.}

\tablenotetext{a}{Approximate absolute magnitude ($H$) or approximate apparent magnitude in a typical optical filter ($V$) of the satellite. These are calculated combining the relative magnitude with the absolute and typical apparent magnitudes of the KBOs from JPL Horizons. These are meant
mostly for illustration purposes and generally have significant uncertainties of $\lesssim$1 magnitude.}
\tablenotetext{b}{Radius estimate in kilometers, listed for illustration purposes only. Quoted radii for the highly ellipsoidal small satellites of Pluto are volumetric means (S. Porter, pers. comm.). Note that these have albedoes of 0.5, somewhat less than assumed for Haumea's moons. For simplicity and ease of inter-comparison, observed moons of Eris and Makemake are given an estimated albedo of 0.7 like the Haumea moons. The actual albedo and size of these moons is not well constrained.}
\tablenotetext{c}{Approximate semi-major axis in units of thousands of kilometers. For upper-limits, this is the approximate range of semi-major axes where this limit applies. The discovery of S/2015 (136472) 1 by \citet{2016arXiv160407461P} within the magnitude and distance ``upper-limit'' quoted by \citet{2008ssbn.book..335B} is easily attributed to the difficulty of detecting moons with small semi-major axes and/or edge-on orbits in single-epoch observations when the actual on-the-sky separation is often small enough to render the moon indistinguishable from the primary \citep{2016arXiv160407461P}. The upper-limits reported here should be understood with that caveat.
}

\tablerefs{
(1) RB09 \citep{2009AJ....137.4766R} \quad
(2) $\S$4, this paper \quad
(3) \citet{2006Natur.439..943W} \quad
(4) \citet{2011IAUC.9221....1S} \quad
(5) \citet{2012IAUC.9253....1S} \quad
(6) \citet{2007Sci...316.1585B} \quad
(7) \citet{2008ssbn.book..335B} \quad
(8) \citet{2016arXiv160407461P}
}
\end{deluxetable*}

\subsection{Distant Satellites}

Satellites with semi-major axes beyond 350000 km may not have been detected in the Program 12243 WFC3 data due to the small field-of-view employed for the subarray observations. Other HST and Keck observations that were not as deep covered a larger area and were also searched for satellites. We estimate that satellites larger than about 40 km in radius (again assuming an albedo similar to Haumea's) would have been detected even several tens of arcseconds away by, e.g., the WFPC2 observations (with a field of view of 162"). Because the motion of sattelites in this region is negligible over the relevant timescales, the shift-and-stack method is not necessary.

Using half the size of Haumea's Hill sphere at perihelion as an estimate of the full region of stable satellites \citep{2008AJ....136.2453S}, the semi-major axis of the most distant stable satellites would be about 4.6 $\times$ $10^6$ km or 124". About half of this volume has been covered down to 40 km in radius.

For comparison, the Program 12243 deep observations covered separations up to about 10" around Haumea or 350000 km. Thus, this limit on very small intermediate-range satellites corresponds to about 0.5\% of the stable region radius.

\section{CONCLUSIONS}

By efficient application of the PK10 method for non-linear shift-and-stack and recovery of known implanted sources, we have strongly limited the possibility of undetected satellites in orbit around Haumea. As Figure \ref{svsb} shows, we detect no satellites larger than $\sim$8 km in radius with separations between 10000 and 350000 km. This same region around Pluto contains Charon and 4 small satellites which, by size, would all have been detected in this search.

Nearer to Haumea, diffraction limits make distinguishing small satellites difficult, but there are dynamical reasons to expect that this region is mostly unpopulated. Further from Haumea, other observations would have detected satellites larger than $\sim$40 km in radius within much of the entire region of possible stable satellites. Significant improvement in the detection limits on smaller satellites would require extensive observations that are unlikely in the foreseeable future until, perhaps, deep observations with the James Webb Space Telescope.

Though Pluto contains multiple small moons and some formation theories \citep[e.g.,][]{2008arXiv0802.2951L} predict them in the Haumea system, we find no additional Haumean moons. Considering upper limits from other studies (summarized in Table 1), Nix/Hydra analogues would have been discovered if present around Makemake and they would be near the detection threshold around Eris.

As the properties of the dwarf planet satellite systems differ significantly, it was not anticipated that Pluto's small satellites would necessarily find counterparts around Haumea, though it seems that Makemake may have a satellite of similar size \citep{2016arXiv160407461P}. Our null result affirms that, for the time being, Pluto is the only known KBO with a retinue of small satellites, though such could have been detected or nearly detected around all four dwarf planets. This implies that the satellite systems may result from somewhat different formation pathways, although all the dwarf planet satellites are probably connected with a collisional formation. Pluto's small satellite system may be connected with Charon since, from a dynamical perspective, the other dwarf planet satellites are more like small moons compared to the near-equal-sized Pluto-Charon binary.

We demonstrate that the non-linear shift-and-stack is a valuable tool for satellite searches. Utilizing the application techniques developed herein, this method can sufficiently capture the nonlinearity of the orbits of fast-moving satellites close to the primary. We have applied this technique to the regime of searching for sub-threshold satellites around Haumea, but it could also be used for other long-observation datasets (PK10). Besides discovery of new moons, it has promise for improving astrometric parameters for known faint moving satellites (e.g., precovery observations of Styx and Kerberos). The tractability of the non-linear shift-and-stack also promotes the possibility of applying this to the general search for KBOs, as originally proposed by PK10. Other applications for improving sensitivity are also possible, e.g. searching for moving exoplanets in direct imaging campaigns \citep{2013ApJ...771...10M}. To facilitate further analyses, all data and source codes used in this project are available upon request.

The sensitivity and tractability of the method presented in this work suggests that, when appropriate, it should be applied to other satellite searches in the solar system. The non-detection of small satellites around Haumea increases our understanding of this intriguing object and contributes to the our understanding of the formation and evolution of multiple KBO systems.

\acknowledgements
We thank Alex Parker, Danielle Hastings, and the anonymous referee for discussions and suggestions that improved the manuscript. DR acknowledges the support of a Harvard Institute for Theory and Computation Fellowship. This work is based on NASA/ESA Hubble Space Telescope Program 12243. Support was provided by NASA through grants HST-GO-12243 from the Space Telescope Science Institute (STScI), which is operated by the Association of Universities for Research in Astronomy, Inc., under NASA contract NAS 5-26555.

\bibliographystyle{apj}
\bibliography{ms}

\begin{thebibliography}{53}
\expandafter\ifx\csname natexlab\endcsname\relax\def\natexlab#1{#1}\fi

\bibitem[{{Benecchi} {et~al.}(2010){Benecchi}, {Noll}, {Grundy}, \&
  {Levison}}]{2010Icar..207..978B}
{Benecchi}, S.~D., {Noll}, K.~S., {Grundy}, W.~M., \& {Levison}, H.~F. 2010,
  \icarus, 207, 978

\bibitem[{{Bernstein} {et~al.}(2004){Bernstein}, {Trilling}, {Allen}, {Brown},
  {Holman}, \& {Malhotra}}]{2004AJ....128.1364B}
{Bernstein}, G.~M., {Trilling}, D.~E., {Allen}, R.~L., {Brown}, M.~E.,
  {Holman}, M., \& {Malhotra}, R. 2004, \aj, 128, 1364

\bibitem[{{Brown}(2008)}]{2008ssbn.book..335B}
{Brown}, M.~E. 2008, {The Largest Kuiper Belt Objects} (The Solar System Beyond
  Neptune), 335--344

\bibitem[{{Brown} {et~al.}(2007){Brown}, {Barkume}, {Ragozzine}, \&
  {Schaller}}]{2007Nature..446..296}
{Brown}, M.~E., {Barkume}, K.~M., {Ragozzine}, D., \& {Schaller}, E.~L. 2007,
  \nat, 446, 294

\bibitem[{{Brown} {et~al.}(2005){Brown}, {Bouchez}, {Rabinowitz}, {Sari},
  {Trujillo}, {van Dam}, {Campbell}, {Chin}, {Hartman}, {Johansson}, {Lafon},
  {Le Mignant}, {Stomski}, {Summers}, \& {Wizinowich}}]{2005ApJ...632L..45B}
{Brown}, M.~E., {Bouchez}, A.~H., {Rabinowitz}, D., {Sari}, R., {Trujillo},
  C.~A., {van Dam}, M., {Campbell}, R., {Chin}, J., {Hartman}, S., {Johansson},
  E., {Lafon}, R., {Le Mignant}, D., {Stomski}, P., {Summers}, D., \&
  {Wizinowich}, P. 2005, \apjl, 632, L45

\bibitem[{{Brown} \& {Schaller}(2007)}]{2007Sci...316.1585B}
{Brown}, M.~E., \& {Schaller}, E.~L. 2007, Science, 316, 1585

\bibitem[{{Brown} {et~al.}(2012){Brown}, {Schaller}, \&
  {Fraser}}]{2012AJ....143..146B}
{Brown}, M.~E., {Schaller}, E.~L., \& {Fraser}, W.~C. 2012, \aj, 143, 146

\bibitem[{{Brown} {et~al.}(2006){Brown}, {van Dam}, {Bouchez}, {Le Mignant},
  {Campbell}, {Chin}, {Conrad}, {Hartman}, {Johansson}, {Lafon}, {Rabinowitz},
  {Stomski}, {Summers}, {Trujillo}, \& {Wizinowich}}]{2006ApJ...639L..43B}
{Brown}, M.~E., {van Dam}, M.~A., {Bouchez}, A.~H., {Le Mignant}, D.,
  {Campbell}, R.~D., {Chin}, J.~C.~Y., {Conrad}, A., {Hartman}, S.~K.,
  {Johansson}, E.~M., {Lafon}, R.~E., {Rabinowitz}, D.~L., {Stomski}, Jr.,
  P.~J., {Summers}, D.~M., {Trujillo}, C.~A., \& {Wizinowich}, P.~L. 2006,
  \apjl, 639, L43

\bibitem[{{Canup}(2011)}]{2011AJ....141...35C}
{Canup}, R.~M. 2011, \aj, 141, 35

\bibitem[{{Canup} {et~al.}(1999){Canup}, {Levison}, \&
  {Stewart}}]{1999AJ....117..603C}
{Canup}, R.~M., {Levison}, H.~F., \& {Stewart}, G.~R. 1999, \aj, 117, 603

\bibitem[{{Carry} {et~al.}(2012){Carry}, {Snodgrass}, {Lacerda}, {Hainaut}, \&
  {Dumas}}]{2012A&A...544A.137C}
{Carry}, B., {Snodgrass}, C., {Lacerda}, P., {Hainaut}, O., \& {Dumas}, C.
  2012, \aap, 544, A137

\bibitem[{{Cheng} {et~al.}(2014{\natexlab{a}}){Cheng}, {Lee}, \&
  {Peale}}]{2014Icar..233..242C}
{Cheng}, W.~H., {Lee}, M.~H., \& {Peale}, S.~J. 2014{\natexlab{a}}, \icarus,
  233, 242

\bibitem[{{Cheng} {et~al.}(2014{\natexlab{b}}){Cheng}, {Peale}, \&
  {Lee}}]{2014arXiv1407.1059C}
{Cheng}, W.~H., {Peale}, S.~J., \& {Lee}, M.~H. 2014{\natexlab{b}}, ArXiv
  e-prints

\bibitem[{{{\'C}uk} {et~al.}(2013){{\'C}uk}, {Ragozzine}, \&
  {Nesvorn{\'y}}}]{2013AJ....146...89C}
{{\'C}uk}, M., {Ragozzine}, D., \& {Nesvorn{\'y}}, D. 2013, \aj, 146, 89

\bibitem[{{Elliot} {et~al.}(2010){Elliot}, {Person}, {Zuluaga}, {Bosh},
  {Adams}, {Brothers}, {Gulbis}, {Levine}, {Lockhart}, {Zangari}, {Babcock},
  {Dupr{\'e}}, {Pasachoff}, {Souza}, {Rosing}, {Secrest}, {Bright}, {Dunham},
  {Sheppard}, {Kakkala}, {Tilleman}, {Berger}, {Briggs}, {Jacobson}, {Valleli},
  {Volz}, {Rapoport}, {Hart}, {Brucker}, {Michel}, {Mattingly},
  {Zambrano-Marin}, {Meyer}, {Wolf}, {Ryan}, {Ryan}, {Morzinski}, {Grigsby},
  {Brimacombe}, {Ragozzine}, {Montano}, \& {Gilmore}}]{2010Natur.465..897E}
{Elliot}, J.~L., {Person}, M.~J., {Zuluaga}, C.~A., {Bosh}, A.~S., {Adams},
  E.~R., {Brothers}, T.~C., {Gulbis}, A.~A.~S., {Levine}, S.~E., {Lockhart},
  M., {Zangari}, A.~M., {Babcock}, B.~A., {Dupr{\'e}}, K., {Pasachoff}, J.~M.,
  {Souza}, S.~P., {Rosing}, W., {Secrest}, N., {Bright}, L., {Dunham}, E.~W.,
  {Sheppard}, S.~S., {Kakkala}, M., {Tilleman}, T., {Berger}, B., {Briggs},
  J.~W., {Jacobson}, G., {Valleli}, P., {Volz}, B., {Rapoport}, S., {Hart}, R.,
  {Brucker}, M., {Michel}, R., {Mattingly}, A., {Zambrano-Marin}, L., {Meyer},
  A.~W., {Wolf}, J., {Ryan}, E.~V., {Ryan}, W.~H., {Morzinski}, K., {Grigsby},
  B., {Brimacombe}, J., {Ragozzine}, D., {Montano}, H.~G., \& {Gilmore}, A.
  2010, \nat, 465, 897

\bibitem[{{Fraser} \& {Brown}(2012)}]{2012ApJ...749...33F}
{Fraser}, W.~C., \& {Brown}, M.~E. 2012, \apj, 749, 33

\bibitem[{{Holman} {et~al.}(2004){Holman}, {Kavelaars}, {Grav}, {Gladman},
  {Fraser}, {Milisavljevic}, {Nicholson}, {Burns}, {Carruba}, {Petit},
  {Rousselot}, {Mousis}, {Marsden}, \& {Jacobson}}]{2004Natur.430..865H}
{Holman}, M.~J., {Kavelaars}, J.~J., {Grav}, T., {Gladman}, B.~J., {Fraser},
  W.~C., {Milisavljevic}, D., {Nicholson}, P.~D., {Burns}, J.~A., {Carruba},
  V., {Petit}, J.-M., {Rousselot}, P., {Mousis}, O., {Marsden}, B.~G., \&
  {Jacobson}, R.~A. 2004, \nat, 430, 865

\bibitem[{{Holman} \& {Wiegert}(1999)}]{1999AJ....117..621H}
{Holman}, M.~J., \& {Wiegert}, P.~A. 1999, \aj, 117, 621

\bibitem[{{Kavelaars} {et~al.}(2004){Kavelaars}, {Holman}, {Grav},
  {Milisavljevic}, {Fraser}, {Gladman}, {Petit}, {Rousselot}, {Mousis}, \&
  {Nicholson}}]{2004Icar..169..474K}
{Kavelaars}, J.~J., {Holman}, M.~J., {Grav}, T., {Milisavljevic}, D., {Fraser},
  W., {Gladman}, B.~J., {Petit}, J.-M., {Rousselot}, P., {Mousis}, O., \&
  {Nicholson}, P.~D. 2004, \icarus, 169, 474

\bibitem[{{Kenyon} \& {Bromley}(2014)}]{2014AJ....147....8K}
{Kenyon}, S.~J., \& {Bromley}, B.~C. 2014, \aj, 147, 8

\bibitem[{{Lacerda}(2009)}]{2009AJ....137.3404L}
{Lacerda}, P. 2009, \aj, 137, 3404

\bibitem[{{Lacerda} {et~al.}(2008){Lacerda}, {Jewitt}, \&
  {Peixinho}}]{2008AJ....135.1749L}
{Lacerda}, P., {Jewitt}, D., \& {Peixinho}, N. 2008, \aj, 135, 1749

\bibitem[{{Leinhardt} {et~al.}(2010){Leinhardt}, {Marcus}, \&
  {Stewart}}]{2010ApJ...714.1789L}
{Leinhardt}, Z.~M., {Marcus}, R.~A., \& {Stewart}, S.~T. 2010, \apj, 714, 1789

\bibitem[{{Levison} {et~al.}(2008){Levison}, {Morbidelli}, {Vokrouhlick{\'y}},
  \& {Bottke}}]{2008AJ....136.1079L}
{Levison}, H.~F., {Morbidelli}, A., {Vokrouhlick{\'y}}, D., \& {Bottke}, W.~F.
  2008, \aj, 136, 1079

\bibitem[{{Lithwick} \& {Wu}(2008)}]{2008arXiv0802.2951L}
{Lithwick}, Y., \& {Wu}, Y. 2008, ArXiv e-prints

\bibitem[{{Lockwood} {et~al.}(2014){Lockwood}, {Brown}, \&
  {Stansberry}}]{2014EM&P..111..127L}
{Lockwood}, A.~C., {Brown}, M.~E., \& {Stansberry}, J. 2014, Earth Moon and
  Planets, 111, 127

\bibitem[{{Lykawka} {et~al.}(2012){Lykawka}, {Horner}, {Mukai}, \&
  {Nakamura}}]{2012MNRAS.421.1331L}
{Lykawka}, P.~S., {Horner}, J., {Mukai}, T., \& {Nakamura}, A.~M. 2012, \mnras,
  421, 1331

\bibitem[{{Males} {et~al.}(2013){Males}, {Skemer}, \&
  {Close}}]{2013ApJ...771...10M}
{Males}, J.~R., {Skemer}, A.~J., \& {Close}, L.~M. 2013, \apj, 771, 10

\bibitem[{{Marcus} {et~al.}(2011){Marcus}, {Ragozzine}, {Murray-Clay}, \&
  {Holman}}]{2011ApJ...733...40M}
{Marcus}, R.~A., {Ragozzine}, D., {Murray-Clay}, R.~A., \& {Holman}, M.~J.
  2011, \apj, 733, 40

\bibitem[{{Markwardt}(2009)}]{2009ASPC..411..251M}
{Markwardt}, C.~B. 2009, in Astronomical Society of the Pacific Conference
  Series, Vol. 411, Astronomical Data Analysis Software and Systems XVIII, ed.
  D.~A. {Bohlender}, D.~{Durand}, \& P.~{Dowler}, 251

\bibitem[{{Noll} {et~al.}(2008){Noll}, {Grundy}, {Chiang}, {Margot}, \&
  {Kern}}]{2008ssbn.book..345N}
{Noll}, K.~S., {Grundy}, W.~M., {Chiang}, E.~I., {Margot}, J.-L., \& {Kern},
  S.~D. 2008, {Binaries in the Kuiper Belt} (The Solar System Beyond Neptune),
  345--363

\bibitem[{{Ortiz} {et~al.}(2012){Ortiz}, {Thirouin}, {Campo Bagatin},
  {Duffard}, {Licandro}, {Richardson}, {Santos-Sanz}, {Morales}, \&
  {Benavidez}}]{2012MNRAS.419.2315O}
{Ortiz}, J.~L., {Thirouin}, A., {Campo Bagatin}, A., {Duffard}, R., {Licandro},
  J., {Richardson}, D.~C., {Santos-Sanz}, P., {Morales}, N., \& {Benavidez},
  P.~G. 2012, \mnras, 419, 2315

\bibitem[{{Parker} {et~al.}(2016){Parker}, {Buie}, {Grundy}, \&
  {Noll}}]{2016arXiv160407461P}
{Parker}, A.~H., {Buie}, M.~W., {Grundy}, W.~M., \& {Noll}, K.~S. 2016, ArXiv
  e-prints

\bibitem[{{Parker} \& {Kavelaars}(2010)}]{2010PASP..122..549P}
{Parker}, A.~H., \& {Kavelaars}, J.~J. 2010, \pasp, 122, 549

\bibitem[{{Pires dos Santos} {et~al.}(2012){Pires dos Santos}, {Morbidelli}, \&
  {Nesvorn{\'y}}}]{2012CeMDA.114..341P}
{Pires dos Santos}, P.~M., {Morbidelli}, A., \& {Nesvorn{\'y}}, D. 2012,
  Celestial Mechanics and Dynamical Astronomy, 114, 341

\bibitem[{{Porter} \& {Stern}(2015)}]{2015arXiv150505933P}
{Porter}, S.~B., \& {Stern}, S.~A. 2015, ArXiv e-prints

\bibitem[{{Rabinowitz} {et~al.}(2006){Rabinowitz}, {Barkume}, {Brown}, {Roe},
  {Schwartz}, {Tourtellotte}, \& {Trujillo}}]{2006ApJ...639.1238R}
{Rabinowitz}, D.~L., {Barkume}, K., {Brown}, M.~E., {Roe}, H., {Schwartz}, M.,
  {Tourtellotte}, S., \& {Trujillo}, C. 2006, \apj, 639, 1238

\bibitem[{{Ragozzine} \& {Brown}(2007)}]{2007AJ....134.2160R}
{Ragozzine}, D., \& {Brown}, M.~E. 2007, \aj, 134, 2160

\bibitem[{{Ragozzine} \& {Brown}(2009)}]{2009AJ....137.4766R}
---. 2009, \aj, 137, 4766

\bibitem[{{Schaller} \& {Brown}(2008)}]{2008ApJ...684L.107S}
{Schaller}, E.~L., \& {Brown}, M.~E. 2008, \apjl, 684, L107

\bibitem[{{Scheeres}(1994)}]{1994Icar..110..225S}
{Scheeres}, D.~J. 1994, Icarus, 110, 225

\bibitem[{{Schlichting} \& {Sari}(2009)}]{2009ApJ...700.1242S}
{Schlichting}, H.~E., \& {Sari}, R. 2009, \apj, 700, 1242

\bibitem[{{Shen} \& {Tremaine}(2008)}]{2008AJ....136.2453S}
{Shen}, Y., \& {Tremaine}, S. 2008, \aj, 136, 2453

\bibitem[{{Showalter} {et~al.}(2013){Showalter}, {de Pater}, {French}, \&
  {Lissauer}}]{2013DPS....4520601S}
{Showalter}, M.~R., {de Pater}, I., {French}, R.~S., \& {Lissauer}, J.~J. 2013,
  in AAS/Division for Planetary Sciences Meeting Abstracts, Vol.~45,
  AAS/Division for Planetary Sciences Meeting Abstracts, 206.01

\bibitem[{{Showalter} \& {Hamilton}(2015)}]{2015Natur.522...45S}
{Showalter}, M.~R., \& {Hamilton}, D.~P. 2015, \nat, 522, 45

\bibitem[{{Showalter} {et~al.}(2011){Showalter}, {Hamilton}, {Stern}, {Weaver},
  {Steffl}, \& {Young}}]{2011IAUC.9221....1S}
{Showalter}, M.~R., {Hamilton}, D.~P., {Stern}, S.~A., {Weaver}, H.~A.,
  {Steffl}, A.~J., \& {Young}, L.~A. 2011, \iaucirc, 9221, 1

\bibitem[{{Showalter} {et~al.}(2012){Showalter}, {Weaver}, {Stern}, {Steffl},
  {Buie}, {Merline}, {Mutchler}, {Soummer}, \& {Throop}}]{2012IAUC.9253....1S}
{Showalter}, M.~R., {Weaver}, H.~A., {Stern}, S.~A., {Steffl}, A.~J., {Buie},
  M.~W., {Merline}, W.~J., {Mutchler}, M.~J., {Soummer}, R., \& {Throop}, H.~B.
  2012, \iaucirc, 9253, 1

\bibitem[{{Snodgrass} {et~al.}(2010){Snodgrass}, {Carry}, {Dumas}, \&
  {Hainaut}}]{2010A&A...511A..72S}
{Snodgrass}, C., {Carry}, B., {Dumas}, C., \& {Hainaut}, O. 2010, \aap, 511,
  A72

\bibitem[{{Stern} {et~al.}(2015){Stern}, {Bagenal}, {Ennico}, {Gladstone},
  {Grundy}, {McKinnon}, {Moore}, {Olkin}, {Spencer}, {Weaver}, {Young},
  {Andert}, {Andrews}, {Banks}, {Bauer}, {Bauman}, {Barnouin}, {Bedini},
  {Beisser}, {Beyer}, {Bhaskaran}, {Binzel}, {Birath}, {Bird}, {Bogan},
  {Bowman}, {Bray}, {Brozovic}, {Bryan}, {Buckley}, {Buie}, {Buratti},
  {Bushman}, {Calloway}, {Carcich}, {Cheng}, {Conard}, {Conrad}, {Cook},
  {Cruikshank}, {Custodio}, {Dalle Ore}, {Deboy}, {Dischner}, {Dumont},
  {Earle}, {Elliott}, {Ercol}, {Ernst}, {Finley}, {Flanigan}, {Fountain},
  {Freeze}, {Greathouse}, {Green}, {Guo}, {Hahn}, {Hamilton}, {Hamilton},
  {Hanley}, {Harch}, {Hart}, {Hersman}, {Hill}, {Hill}, {Hinson}, {Holdridge},
  {Horanyi}, {Howard}, {Howett}, {Jackman}, {Jacobson}, {Jennings}, {Kammer},
  {Kang}, {Kaufmann}, {Kollmann}, {Krimigis}, {Kusnierkiewicz}, {Lauer}, {Lee},
  {Lindstrom}, {Linscott}, {Lisse}, {Lunsford}, {Mallder}, {Martin}, {McComas},
  {McNutt}, {Mehoke}, {Mehoke}, {Melin}, {Mutchler}, {Nelson}, {Nimmo},
  {Nunez}, {Ocampo}, {Owen}, {Paetzold}, {Page}, {Parker}, {Parker},
  {Pelletier}, {Peterson}, {Pinkine}, {Piquette}, {Porter}, {Protopapa},
  {Redfern}, {Reitsema}, {Reuter}, {Roberts}, {Robbins}, {Rogers}, {Rose},
  {Runyon}, {Retherford}, {Ryschkewitsch}, {Schenk}, {Schindhelm}, {Sepan},
  {Showalter}, {Singer}, {Soluri}, {Stanbridge}, {Steffl}, {Strobel}, {Stryk},
  {Summers}, {Szalay}, {Tapley}, {Taylor}, {Taylor}, {Throop}, {Tsang},
  {Tyler}, {Umurhan}, {Verbiscer}, {Versteeg}, {Vincent}, {Webbert}, {Weidner},
  {Weigle}, {White}, {Whittenburg}, {Williams}, {Williams}, {Williams},
  {Woods}, {Zangari}, \& {Zirnstein}}]{2015Sci...350.1815S}
{Stern}, S.~A., {Bagenal}, F., {Ennico}, K., {Gladstone}, G.~R., {Grundy},
  W.~M., {McKinnon}, W.~B., {Moore}, J.~M., {Olkin}, C.~B., {Spencer}, J.~R.,
  {Weaver}, H.~A., {Young}, L.~A., {Andert}, T., {Andrews}, J., {Banks}, M.,
  {Bauer}, B., {Bauman}, J., {Barnouin}, O.~S., {Bedini}, P., {Beisser}, K.,
  {Beyer}, R.~A., {Bhaskaran}, S., {Binzel}, R.~P., {Birath}, E., {Bird}, M.,
  {Bogan}, D.~J., {Bowman}, A., {Bray}, V.~J., {Brozovic}, M., {Bryan}, C.,
  {Buckley}, M.~R., {Buie}, M.~W., {Buratti}, B.~J., {Bushman}, S.~S.,
  {Calloway}, A., {Carcich}, B., {Cheng}, A.~F., {Conard}, S., {Conrad}, C.~A.,
  {Cook}, J.~C., {Cruikshank}, D.~P., {Custodio}, O.~S., {Dalle Ore}, C.~M.,
  {Deboy}, C., {Dischner}, Z.~J.~B., {Dumont}, P., {Earle}, A.~M., {Elliott},
  H.~A., {Ercol}, J., {Ernst}, C.~M., {Finley}, T., {Flanigan}, S.~H.,
  {Fountain}, G., {Freeze}, M.~J., {Greathouse}, T., {Green}, J.~L., {Guo}, Y.,
  {Hahn}, M., {Hamilton}, D.~P., {Hamilton}, S.~A., {Hanley}, J., {Harch}, A.,
  {Hart}, H.~M., {Hersman}, C.~B., {Hill}, A., {Hill}, M.~E., {Hinson}, D.~P.,
  {Holdridge}, M.~E., {Horanyi}, M., {Howard}, A.~D., {Howett}, C.~J.~A.,
  {Jackman}, C., {Jacobson}, R.~A., {Jennings}, D.~E., {Kammer}, J.~A., {Kang},
  H.~K., {Kaufmann}, D.~E., {Kollmann}, P., {Krimigis}, S.~M.,
  {Kusnierkiewicz}, D., {Lauer}, T.~R., {Lee}, J.~E., {Lindstrom}, K.~L.,
  {Linscott}, I.~R., {Lisse}, C.~M., {Lunsford}, A.~W., {Mallder}, V.~A.,
  {Martin}, N., {McComas}, D.~J., {McNutt}, R.~L., {Mehoke}, D., {Mehoke}, T.,
  {Melin}, E.~D., {Mutchler}, M., {Nelson}, D., {Nimmo}, F., {Nunez}, J.~I.,
  {Ocampo}, A., {Owen}, W.~M., {Paetzold}, M., {Page}, B., {Parker}, A.~H.,
  {Parker}, J.~W., {Pelletier}, F., {Peterson}, J., {Pinkine}, N., {Piquette},
  M., {Porter}, S.~B., {Protopapa}, S., {Redfern}, J., {Reitsema}, H.~J.,
  {Reuter}, D.~C., {Roberts}, J.~H., {Robbins}, S.~J., {Rogers}, G., {Rose},
  D., {Runyon}, K., {Retherford}, K.~D., {Ryschkewitsch}, M.~G., {Schenk}, P.,
  {Schindhelm}, E., {Sepan}, B., {Showalter}, M.~R., {Singer}, K.~N., {Soluri},
  M., {Stanbridge}, D., {Steffl}, A.~J., {Strobel}, D.~F., {Stryk}, T.,
  {Summers}, M.~E., {Szalay}, J.~R., {Tapley}, M., {Taylor}, A., {Taylor}, H.,
  {Throop}, H.~B., {Tsang}, C.~C.~C., {Tyler}, G.~L., {Umurhan}, O.~M.,
  {Verbiscer}, A.~J., {Versteeg}, M.~H., {Vincent}, M., {Webbert}, R.,
  {Weidner}, S., {Weigle}, G.~E., {White}, O.~L., {Whittenburg}, K.,
  {Williams}, B.~G., {Williams}, K., {Williams}, S., {Woods}, W.~W., {Zangari},
  A.~M., \& {Zirnstein}, E. 2015, Science, 350, aad1815

\bibitem[{{Volk} \& {Malhotra}(2012)}]{2012Icar..221..106V}
{Volk}, K., \& {Malhotra}, R. 2012, \icarus, 221, 106

\bibitem[{{Ward} \& {Canup}(2006)}]{2006Sci...313.1107W}
{Ward}, W.~R., \& {Canup}, R.~M. 2006, Science, 313, 1107

\bibitem[{{Weaver} {et~al.}(2006){Weaver}, {Stern}, {Mutchler}, {Steffl},
  {Buie}, {Merline}, {Spencer}, {Young}, \& {Young}}]{2006Natur.439..943W}
{Weaver}, H.~A., {Stern}, S.~A., {Mutchler}, M.~J., {Steffl}, A.~J., {Buie},
  M.~W., {Merline}, W.~J., {Spencer}, J.~R., {Young}, E.~F., \& {Young}, L.~A.
  2006, \nat, 439, 943

\bibitem[{{Youdin} {et~al.}(2012){Youdin}, {Kratter}, \&
  {Kenyon}}]{2012ApJ...755...17Y}
{Youdin}, A.~N., {Kratter}, K.~M., \& {Kenyon}, S.~J. 2012, \apj, 755, 17

\end{thebibliography}

\end{document}